\documentclass[twocolumn,aps,superscriptaddress,showpacs]{revtex4}
\usepackage{amsmath,bm}
\usepackage{graphicx}

\begin{document}

\title{Nuclear matter symmetry energy and the neutron skin thickness of
heavy nuclei}
\author{Lie-Wen Chen}
\affiliation{Institute of Theoretical Physics, Shanghai Jiao Tong University, Shanghai
200240, China}
\affiliation{Center of Theoretical Nuclear Physics, National Laboratory of Heavy Ion
Accelerator, Lanzhou 730000, China}
\author{Che Ming Ko}
\affiliation{Cyclotron Institute and Physics Department, Texas A\&M University, College
Station, Texas 77843-3366}
\author{Bao-An Li}
\affiliation{Department of Chemistry and Physics, P.O. Box 419, Arkansas State
University, State University, Arkansas 72467-0419}
\date{\today}

\begin{abstract}
Correlations between the thickness of the neutron skin in finite nuclei and
the nuclear matter symmetry energy are studied in the Skyrme Hartree-Fock
model. From the most recent analysis of the isospin diffusion data in
heavy-ion collisions based on an isospin- and momentum-dependent transport
model with in-medium nucleon-nucleon cross sections, a value of $L=88\pm 25$
MeV for the slope of the nuclear symmetry energy at saturation density is
extracted, and this imposes stringent constraints on both the parameters in
the Skyrme effective interactions and the neutron skin thickness of heavy
nuclei. Predicted thickness of the neutron skin is $0.22\pm 0.04$ fm for $%
^{208}$Pb, $0.29\pm 0.04$ fm for $^{132}$Sn, and $0.22\pm 0.04$ fm for $%
^{124}$Sn.
\end{abstract}

\maketitle

\section{Introduction}

The study of the equation of state (EOS) of isospin asymmetric nuclear
matter, especially the nuclear symmetry energy, is currently an active field
of research in nuclear physics \cite%
{ireview98,ibook,bom,diep03,pawel02,lat04,baran05,steiner05}. Although the
nuclear symmetry energy at normal nuclear matter density is known to be
around $30$ \textrm{MeV} from the empirical liquid-drop mass formula \cite%
{myers,pomorski}, its values at other densities, especially at supranormal
densities, are poorly known \cite{ireview98,ibook}. Advances in radioactive
nuclear beam facilities provide, however, the possibility to pin down the
density dependence of the nuclear symmetry energy in heavy ion collisions
induced by these nuclei \cite%
{ireview98,ibook,baran05,li97,muller95,fra1,fra2,xu00,tan01,bar02,betty,lis,li00,npa01,li02,chen03,ono03,liu03,chen04,li04a,shi03,li04prc,rizzo04,liyong05,lizx05a,lizx05b,tian05}%
. Indeed, significant progress has recently been made in extracting the
information on the density dependence of nuclear symmetry energy from the
isospin diffusion data in heavy-ion collisions from the NSCL/MSU \cite%
{tsang04,chen05,li05}. Using an isospin- and momentum-dependent IBUU04
transport model with in-medium nucleon-nucleon (NN) cross sections, the
isospin diffusion data were found to be consistent with a relatively soft
nuclear symmetry energy at subnormal density \cite{li05}.

Information on the density dependence of the nuclear symmetry energy can in
principle also be obtained from the thickness of the neutron skin in heavy
nuclei as the latter is strongly correlated with the slope $L$ of the
nuclear matter symmetry energy at saturation density \cite%
{brown00,hor01,typel01,furn02,kara02,diep03}. Because of the large
uncertainties in measured neutron skin thickness of heavy nuclei, this has
not been possible. Instead, studies have been carried out to use the
extracted nuclear symmetry energy from the isospin diffusion data to
constrain the neutron skin thickness of heavy nuclei \cite{steiner05b,li05}.
Using the Hartree-Fock approximation with parameters fitted to the
phenomenological EOS that was used in the IBUU04 transport model to describe
the isospin diffusion data from the NSCL/MSU, it was found that a neutron
skin thickness of less than $0.15$ fm \cite{steiner05b,li05} for $^{208}$Pb
was incompatible with the isospin diffusion data.

In the present work, we study more systematically the correlation between
the density dependence of the nuclear symmetry energy and the thickness of
the neutron skin in a number of nuclei within the framework of the Skyrme
Hartree-Fock model. Extracting the values of $L$ from the most recently
determined density dependence of the nuclear symmetry energy from the
isospin diffusion data in heavy-ion collisions, we obtain stringent
constraints on the neutron skin thickness of the nuclei $^{208}$Pb, $^{132}$%
Sn, and $^{124}$Sn. The extracted value of $L$ also limits the allowed
parameter sets for the Skyrme interaction.

\section{Nuclear symmetry energy and the Skyrme interactions}

The nuclear symmetry energy $E_{\text{sym}}(\rho )$ at nuclear density $\rho$
can be expanded around the nuclear matter saturation density $\rho _{0}$ as 
\begin{equation}
E_{\text{sym}}(\rho )=E_{\text{sym}}(\rho _{0})+\frac{L}{3}\left( \frac{\rho
-\rho _{0}}{\rho _{0}}\right) +\frac{K_{\text{sym}}}{18}\left( \frac{\rho
-\rho _{0}}{\rho _{0}}\right) ^{2},  \label{EsymLK}
\end{equation}%
where $L$ and $K_{\text{sym}}$ are the slope and curvature of the nuclear
symmetry energy at $\rho _{0}$, i.e., 
\begin{eqnarray}
L &=&3\rho _{0}\frac{\partial E_{\text{sym}}(\rho )}{\partial \rho }|_{\rho
=\rho _{0}},  \label{L} \\
K_{\text{sym}} &=&9\rho _{0}^{2}\frac{\partial ^{2}E_{\text{sym}}(\rho )}{%
\partial ^{2}\rho }|_{\rho =\rho _{0}}.  \label{Ksym}
\end{eqnarray}%
The $L$ and $K_{\text{sym}}$ characterize the density dependence of the
nuclear symmetry energy around normal nuclear matter density, and thus
provide important information on the properties of nuclear symmetry energy
at both high and low densities.

In the standard Skyrme Hartree-Fock model \cite%
{brack85,fried86,brown98,clw99,stone03,brown00}, the interaction is taken to
have a zero-range, density- and momentum-dependent form and the Skyrme
interaction parameters are chosen to fit the binding energies and charge
radii of a large number of nuclei in the periodic table. For infinite
nuclear matter, the nuclear symmetry energy from the Skyrme interaction can
be expressed as \cite{clw99,stone03} 
\begin{eqnarray}
E_{\text{sym}}(\rho ) &=&\frac{1}{3}\frac{\hbar ^{2}}{2m}\left( \frac{3\pi
^{2}}{2}\right) ^{2/3}\rho ^{2/3}  \notag \\
&&-\frac{1}{8}t_{0}(2x_{0}+1)\rho -\frac{1}{48}t_{3}(2x_{3}+1)\rho ^{\sigma
+1}  \notag \\
&&+\frac{1}{24}\left( \frac{3\pi ^{2}}{2}\right) ^{2/3}\left[
-3t_{1}x_{1}\right.  \notag \\
&&+\left. \left( 4+5x_{2}\right) t_{2}\right] \rho ^{5/3},  \label{EsymSky}
\end{eqnarray}%
where the $\sigma $, $t_{0}-t_{3}$, and $x_{0}-x_{3}$ are Skyrme interaction
parameters.

\begin{figure}[tbp]
\includegraphics[scale=0.9]{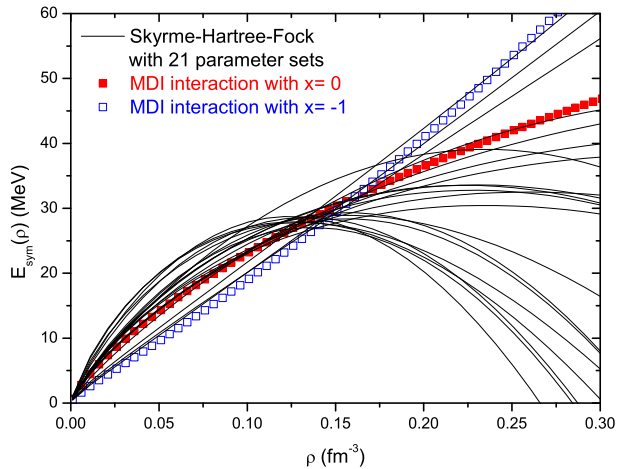}
\caption{{\protect\small (Color online) Density dependence of the nuclear
symmetry energy }$E_{\text{sym}}(\protect\rho )${\protect\small \ for 21
sets of Skyrme interaction parameters. The results from the MDI interaction
with }$x=-1${\protect\small \ (open squares) and }$0${\protect\small \
(solid squares) are also shown.}}
\label{SymDen}
\end{figure}

Fig. \ref{SymDen} displays the density dependence of $E_{\text{sym}}(\rho )$
for $21$ sets of Skyrme interaction parameters, i.e., \textrm{SKM}, \textrm{%
SKM}$^{\ast }$, $\mathrm{RATP}$, \textrm{SI}, \textrm{SII}, $\mathrm{SIII}$, 
\textrm{SIV}, \textrm{SV}, \textrm{SVI}, \textrm{E}, \textrm{E}$_{\sigma }$, 
\textrm{G}$_{\sigma }$, $\mathrm{R}_{\sigma }$, $\mathrm{Z}$, $\mathrm{Z}%
_{\sigma }$, $\mathrm{Z}_{\sigma }^{\ast }$, $\mathrm{T}$, $\mathrm{T3}$, $%
\mathrm{SkX}$, $\mathrm{SkXce}$, and $\mathrm{SkXm}$. Values of the
parameters in these Skyrme interactions can be found in Refs. \cite%
{brack85,fried86,brown98}. For comparison, we also show in Fig. \ref{SymDen}
results from the phenomenological parametrization of the momentum-dependent
nuclear mean-field potential based on the Gogny effective interaction \cite%
{das03}, i.e., the MDI interactions with $x=-1$ (open squares) and $0$
(solid squares), where different $x$ values correspond to different density
dependence of the nuclear symmetry energy but keep other properties of the
nuclear EOS the same \cite{chen05}. From comparing the isospin diffusion
data from NSCL/MSU using the IBUU04 transport model with in-medium NN cross
sections, these interactions are recently shown to give, respectively, the
upper and lower bounds for the stiffness of the nuclear symmetry energy \cite%
{li05}. It is seen from Fig. \ref{SymDen} that the density dependence of the
symmetry energy varies drastically among different interactions. Although
the values of $E_{\text{sym}}(\rho _{0})$ are all in the range of $26$-$35$
MeV, the values of $L$ and $K_{\text{sym}}$ are in the range of $-50$-$100$
MeV and $-700$-$50$ MeV, respectively.

\section{Neutron skin thickness of finite nuclei and the slope of nuclear
symmetry energy at saturation density}

\begin{figure}[tbp]
\includegraphics[scale=0.9]{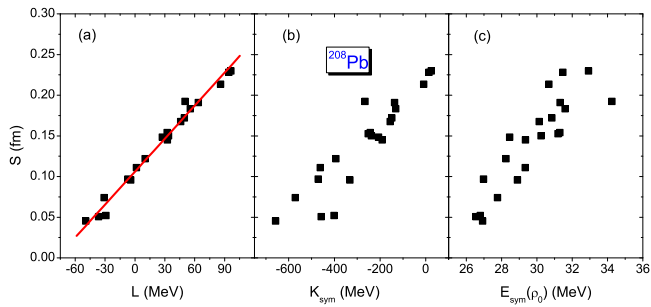}
\caption{{\protect\small (Color online) {Neutron skin thickness $S$ of $%
^{208}$Pb as a function of (a) $L$, (b) $K_{\text{sym}}$, and (c) $E_{\text{%
sym}}(\protect\rho _{0})$ for 21 sets of Skyrme interaction parameters. }The
line in panel (a) represents a linear fit.}}
\label{SPb208}
\end{figure}

The neutron skin thickness $S$ of a nucleus is defined as the difference
between the root-mean-square radii $\sqrt{\left\langle r_{n}\right\rangle }$
of neutrons and $\sqrt{\left\langle r_{p}\right\rangle }$ of protons, i.e., 
\begin{equation}
S=\sqrt{\left\langle r_{n}^{2}\right\rangle }-\sqrt{\left\langle
r_{p}^{2}\right\rangle }.  \label{S}
\end{equation}%
It has been known that $S$ is sensitive to the density dependence of the
nuclear symmetry energy, particularly the slope parameter $L$ at the normal
nuclear matter density \cite{brown00,hor01,typel01,furn02,kara02,diep03}.
Using above $21$ sets of Skyrme interaction parameters, we have evaluated
the neutron skin thickness of several nuclei. In Figs. \ref{SPb208}(a), (b)
and (c), we show, respectively, the correlations between the neutron skin
thickness of $^{208}$Pb with $L$, $K_{\text{sym}}$, and $E_{\text{sym}}(\rho
_{0})$. It is seen from Fig. \ref{SPb208}(a) that there exists an
approximate linear correlation between $S$ and $L$. The correlations of $S$
with $K_{\text{sym}}$ and $E_{\text{sym}}(\rho _{0})$ are less strong and
even exhibit some irregular behavior. The solid line in Fig. \ref{SPb208}(a)
is a linear fit to the correlation between $S$ and $L$ and is given by the
following expression: 
\begin{eqnarray}
S(^{\text{208}}\text{Pb)} &=&(0.1066\pm 0.0019)  \notag \\
&&+(0.00133\pm 3.76\times 10^{-5})\times L,  \label{SLPb208a}
\end{eqnarray}%
or 
\begin{eqnarray}
L &=&(-78.5\pm 3.2)  \notag \\
&&+(740.4\pm 20.9)\times S(^{\text{208}}\text{Pb)},  \label{SLPb208b}
\end{eqnarray}%
where the units of $L$ and $S$ are \textrm{MeV} and \textrm{fm},
respectively. Therefore, if the value for either $S(^{\text{208}}$Pb) or $L$
is known, the value for the other can be determined. 

\begin{figure}[tbp]
\includegraphics[scale=0.9]{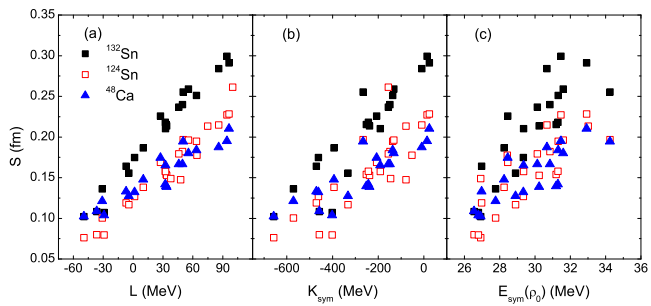}
\caption{{\protect\small (Color online) {Same as Fig. 2 but for nuclei $%
^{132}$Sn (Solid squares), $^{124}$Sn (Open squares) and $^{48}$Ca
(Triangles).}}}
\label{SSnCa}
\end{figure}

It is of interest to see if there are also correlations between the neutron
skin thickness of other neutron-rich nuclei and the nuclear symmetry energy.
Fig. \ref{SSnCa} shows the same correlations as in Fig. \ref{SPb208} but for
the neutron-rich nuclei $^{132}$Sn, $^{124}$Sn, and $^{48}$Ca. For the heavy 
$^{132}$Sn and $^{124}$Sn, we obtain a similar conclusion as for $^{208}$Pb,
namely, $S$ exhibits an approximate linear correlation with $L$ but weaker
correlations with $K_{\text{sym}}$ and $E_{\text{sym}}(\rho _{0})$. For the
lighter $^{48}$Ca, on the other hand, all the correlations become weaker
than those of heavier nuclei. Therefore, the neutron skin thickness of heavy
nuclei is better correlated with the density dependence of the nuclear
symmetry energy. As in Eq. (\ref{SLPb208a}) and (\ref{SLPb208b}), a linear
fit to the correlation between $S$ and $L$ can also be obtained for $^{132}$%
Sn and $^{124}$Sn, and the corresponding expressions are 
\begin{eqnarray}
S(^{\text{132}}\text{Sn)} &=&(0.1694\pm 0.0025)  \notag \\
&&+(0.0014\pm 5.12\times 10^{-5})\times L,  \label{SLSn132a}
\end{eqnarray}%
\begin{eqnarray}
L &=&(-117.1\pm 5.4)  \notag \\
&&+(695.1\pm 25.3)\times S(^{\text{132}}\text{Sn)},  \label{SLSn132b}
\end{eqnarray}%
and 
\begin{eqnarray}
S(^{\text{124}}\text{Sn)} &=&(0.1255\pm 0.0020)  \notag \\
&&+(0.0011\pm 4.05\times 10^{-5})\times L,  \label{SLSn124a}
\end{eqnarray}%
\begin{eqnarray}
L &=&(-110.1\pm 5.2)  \notag \\
&&+(882.6\pm 32.3)\times S(^{\text{124}}\text{Sn)},  \label{SLSn124b}
\end{eqnarray}%

Similar linear relations between $S$ and $L$ are also expected for other
heavy nuclei. This is not surprising as detailed discussions in Refs. \cite%
{brown00,hor01,typel01,furn02,kara02,diep03} have shown that the thickness
of the neutron skin in heavy nuclei is determined by the pressure difference
between neutrons and protons, which is proportional to the parameter $L$. 
\begin{table}[tbp]
\caption{{\protect\small {Linear correlation coefficients $C_{l}$ of $S$
with $L$, $K_{\text{sym}}$ and $E_{\text{sym}}(\protect\rho _{0})$ for $%
^{208}$Pb,\ $^{132}$Sn, $^{124}$Sn, and $^{48}$Ca from 21 sets of Skyrme
interaction parameters.}}}
\label{Corr}%
\begin{tabular}{ccccc}
\hline\hline
$C_{l}$ $(\%)$ & \quad $^{208}$Pb\quad & $\quad ^{132}$Sn \quad & $^{124}$Sn
& $^{48}$Ca \\ \hline
$S$-$L$ & $99.25$ & $98.76$ & $98.75$ & $93.66$ \\ 
$S$-$K_{\text{sym}}$ & $92.26$ & $92.06$ & $92.22$ & $86.99$ \\ 
$S$-$E_{\text{sym}}$ & $87.89$ & $85.74$ & $85.77$ & $81.01$ \\ \hline\hline
\end{tabular}%
\end{table}

To give a quantitative estimate of above discussed correlations, we define
the following linear correlation coefficient $C_{l}$: 
\begin{equation}
C_{l}=\sqrt{1-q/t},
\end{equation}%
where%
\begin{eqnarray}
q &=&\underset{i=1}{\overset{n}{\sum }}[y_{i}-(A+Bx_{i})]^{2}, \\
t &=&\underset{i=1}{\overset{n}{\sum }}(y_{i}-\overline{y}),~~~\overline{y}=%
\underset{i=1}{\overset{n}{\sum }}y_{i}/n.
\end{eqnarray}%
In the above, $A$ and $B$ are the linear regression coefficients, $(x_{i}$, $%
y_{i})$ are the sample points, and $n$ is the number of sample points. The
linear correlation coefficient $C_{l}$ measures the degree of linear
correlation, and $C_{l}=1$ corresponds to an ideal linear correlation. Table %
\ref{Corr} gives the linear correlation coefficient $C_{l}$ for the
correlation of $S$ with $L$, $K_{\text{sym}}$ and $E_{\text{sym}}(\rho _{0})$
for $^{208}$Pb, $^{132}$Sn, $^{124}$Sn, and $^{48}$Ca shown in Figs. \ref%
{SPb208} and \ref{SSnCa} for different Skyrme interactions. It is seen that
these correlations become weaker with decreasing nucleus mass, and a strong
linear correlation only exists between the $S$ and $L$ for the heavier
nuclei $^{208}$Pb, $^{132}$Sn, and $^{124}$Sn. Therefore, the neutron skin
thickness of these nuclei can be extracted once the slope parameter $L$ of
the nuclear symmetry energy at saturation density is known.

\section{Constraints from isospin diffusion data in heavy ion collisions}

Experimentally, the degree of isospin diffusion between the projectile
nucleus $A$ and the target nucleus $B$ can be studied via the quantity $%
R_{i} $ \cite{rami,tsang04}, 
\begin{equation}
R_{i}=\frac{2X^{A+B}-X^{A+A}-X^{B+B}}{X^{A+A}-X^{B+B}},  \label{Ri}
\end{equation}%
where $X$ is any isospin-sensitive observable. By construction, the value of 
$R_{i}$ is $1~(-1)$ for symmetric $A+A~(B+B)$ reaction. If isospin
equilibrium is reached during the collision as a result of isospin
diffusion, the value of $R_{i}$ is about zero. In the NSCL/MSU experiments
with $A=$ $^{124}$Sn and $B=$ $^{112}$Sn at a beam energy of $50$
MeV/nucleon and an impact parameter about $6$ fm, the isospin asymmetry of
the projectile-like residue was used as the isospin tracer $X$ \cite{tsang04}%
. Using an isospin- and momentum-dependent IBUU04 transport model with
free-space experimental NN cross sections, the dependence of $R_{i}$ on the
nuclear symmetry energy was studied from the average isospin asymmetry of
the projectile-like residue that was calculated from nucleons with local
densities higher than $\rho _{0}/20$ and velocities larger than $1/2$ the
beam velocity in the center-of-mass frame \cite{chen05}. Comparing the
theoretical results with the experimental data has allowed us to extract a
nuclear symmetry energy of $E_{\text{sym}}(\rho )\approx 31.6(\rho /\rho
_{0})^{1.05}$. Including also medium-dependent NN cross sections, which are
important for isospin-dependent observables \cite{li05,li05a}, the isospin
diffusion data leads to an even softer nuclear symmetry energy of $E_{\text{%
sym}}(\rho )\approx 31.6(\rho /\rho _{0})^{\gamma }$ with $\gamma \approx
0.7 $ \cite{li05}.

\begin{figure}[tbp]
\includegraphics[scale=0.95]{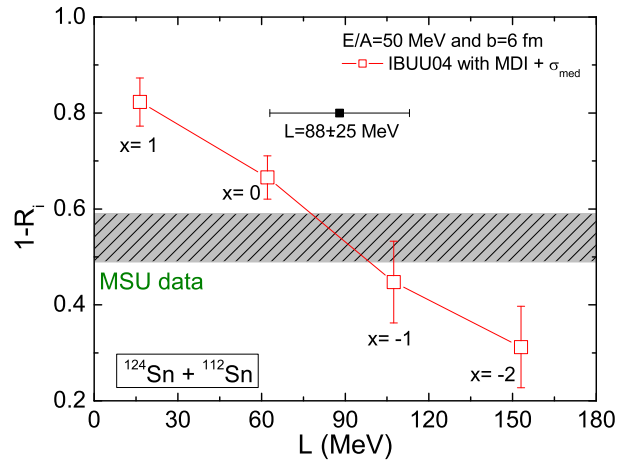}
\caption{{\protect\small (Color online) Degree of the isospin diffusion }$%
1-R_{i}${\protect\small \ as a function of }$L${\protect\small \ using the
MDI interaction with }$x=-2${\protect\small , }$-1${\protect\small , }$0$%
{\protect\small , and }$1${\protect\small . The shaded band indicates the
data from NSCL/MSU \protect\cite{tsang04}. The solid square with error bar
represents }$L=88\pm 25${\protect\small \ MeV.}}
\label{RiL}
\end{figure}

In Fig. \ref{RiL}, we show the results from the IBUU04 transport model with
in-medium NN cross sections, that are consistent with the mean-field
potential obtained with the MDI interactions used in the model, for the
degree of the isospin diffusion $1-R_{i}$ as a function of $L$. The shaded
band in Fig. \ref{RiL} indicates the data from NSCL/MSU \cite{tsang04}. It
is seen that the strength of isospin diffusion $1-R_{i}$ decreases
monotonically with decreasing value of $x$ or increasing value of $L$. This
is expected as the parameter $L$ reflects the difference in the pressures on
neutrons and protons. From comparison of the theoretical results with the
data, we can clearly exclude the MDI interaction with $x=1$ and $x=-2$ as
they give either too large or too small a value for $1-R_{i}$ compared to
that of data. The range of $x$ or $L$ values that give values of $1-R_{i}$
falling within the band of experimental values could in principle be
determined in our model by detailed calculations. Instead, we determine this
schematically by using the results from the four $x$ values. For the
centroid value of $L$, it is obtained from the interception of the line
connecting the theoretical results at $x=-1$ and $0$ with the central value
of $1-R_{i}$ data in Fig. \ref{RiL}, i.e., $L=88$ MeV. The upper limit of $%
L=113$ MeV is estimated from the interception of the line connecting the
upper error bars of the theoretical results at $x=-1$ and $-2$ with the
lower limit of the data band of $1-R_{i}$. Similarly, the lower limit of $%
L=65$ MeV is estimated from the interception of the line connecting the
lower error bars of the theoretical results at $x=0$ and $-1$ with the upper
limit of the data band of $1-R_{i}$. This leads to an extracted value of $%
L=88\pm 25$ MeV as shown by the solid square with error bar in Fig. \ref{RiL}%
.

The extracted value of $L=88\pm 25$ MeV gives a rather stringent constraint
on the density dependence of the nuclear symmetry energy and thus puts
strong constraints on the nuclear effective interactions as well. For the
Skyrme effective interactions shown in Fig. \ref{SymDen}, for instance, all
of those lie beyond $x=0$ and $x=-1$ in the sub-saturation region are not
consistent with the extracted value of $L$. Actually, we note that only $4$
sets of Skyrme interactions, i.e., $\mathrm{SIV}$, $\mathrm{SV}$, $\mathrm{G}%
_{\sigma }$, and $\mathrm{R}_{\sigma }$, in the $21$ sets of Skyrme
interactions considered here have nuclear symmetry energies that are
consistent with the extracted $L$ value.

The extracted $L$ value also allows us to determine from Eqs. (\ref{SLPb208a}%
), (\ref{SLSn132a}), and (\ref{SLSn124a}), respectively, a neutron skin
thickness of $0.22\pm 0.04$ fm for $^{208}$Pb, $0.29\pm 0.04$ fm for $^{132}$%
Sn, and $0.22\pm 0.04$ fm for $^{124}$Sn. Experimentally, great efforts were
devoted to measure the thickness of the neutron skin in heavy nuclei \cite%
{sta94,clark03}, and a recent review can be found in Ref. \cite{kras04}. The
data for the neutron skin thickness of $^{208}$Pb indicate a large
uncertainty, i.e., $0.1$-$0.28$ fm. Our results for the neutron skin
thickness of $^{208}$Pb are thus consistent with present data but give a
much stronger constraint. A large uncertainty is also found experimentally
in the neutron skin thickness of $^{124}$Sn, i.e., its value varies from $%
0.1 $ fm to $0.3$ fm depending on the experimental method. The proposed
experiment of parity-violating electron scattering from $^{208}$Pb at the
Jefferson Laboratory is expected to give another independent and more
accurate measurement of its neutron skin thickness (within $0.05$ fm), thus
providing improved constraints on the density dependence of the nuclear
symmetry energy \cite{hor01b,jeff00}.

Most recently, an accurately calibrated relativistic parametrization based
on the relativistic mean-field theory has been introduced to study the
neutron skin thickness of finite nuclei \cite{piek05}. This parametrization
can describe simultaneously the ground state properties of finite nuclei and
their monopole and dipole resonances. Using this parametrization, the
authors predicted a neutron skin thickness of $0.21$ fm in $^{208}$Pb, $0.27$
fm in $^{132}$Sn, and $0.19$ fm in $^{124}$ Sn \cite{piek05,piek}. These
predictions are in surprisingly good agreement with our results constrained
by the isospin diffusion data in heavy-ion collisions.

\section{Summary}

In summary, we have studied the correlation between the neutron skin
thickness of finite nuclei and the nuclear symmetry energy within the
framework of the Skyrme Hartree-Fock model. As in previous studies, we have
found a strong linear correlation between the neutron skin thickness of
heavy nuclei and the slope $L$ of the nuclear matter symmetry energy at
saturation density. This correlation provides stringent constraints on both
the density dependence of the nuclear symmetry energy and the thickness of
the neutron skin in heavy nuclei. From the most recent analysis of the
isospin diffusion data in heavy-ion collisions using an isospin- and
momentum-dependent transport model with in-medium NN cross sections, the
value $L=88\pm 25$ MeV has been extracted. The relatively constrained value
for the slope of the nuclear matter symmetry energy imposes strong
constraints on the parameters in the Skyrme effective interactions and leads
to predicted neutron skin thickness of $0.22\pm 0.04$ fm for $^{208}$Pb, $%
0.29\pm 0.04$ fm for $^{132}$Sn, and $0.22\pm 0.04$ fm for $^{124}$Sn.

\begin{acknowledgments}
The work was supported in part by the National Natural Science Foundation of
China under Grant No. 10105008 and 10575071 (LWC), the US National Science
Foundation under Grant No. PHY-0457265 and the Welch Foundation under Grant
No. A-1358 (CMK), as well as the US National Science Foundation under Grant
No. PHY-0354572 and PHY-0456890 and the NASA-Arkansas Space Grants
Consortium Award ASU15154 (BAL).
\end{acknowledgments}

\end{document}